# WEB SERVICES SYNCHRONIZATION HEALTH CARE APPLICATION

Hela Limam[1] and Jalel Akaichi[2]

[1, 2] Department of Computer Sciences, ISG,SOIE,Tunis, Tunisia

[1] hela.limam@isg.rnu.tn
[2] jalel.akaichi@isg.rnu.tn

## ABSTRACT

*This With the advance of Web Services technologies and the emergence of Web Services into the information space, tremendous opportunities for empowering users and organizations appear in various application domains including electronic commerce, travel, intelligence information gathering and analysis, health care, digital government, etc. In fact, Web services appear to be s solution for integrating distributed, autonomous and heterogeneous information sources. However, as Web services evolve in a dynamic environment which is the Internet many changes can occur and affect them. A Web service is affected when one or more of its associated information sources is affected by schema changes. Changes can alter the information sources contents but also their schemas which may render Web services partially or totally undefined. In this paper, we propose a solution for integrating information sources into Web services. Then we tackle the Web service synchronization problem by substituting the affected information sources. Our work is illustrated with a healthcare case study.*

## 1. INTRODUCTION

The incredible growth of the information space and the increasing number of available information sources are factors which arise a growing interest for integrating information sources into Web services in order to enhance collaboration and knowledge sharing between enterprises . The emergence of Web services as a model for integrating heterogeneous web information has opened up new possibilities of interaction and offered more potential for interoperability. However, the organization into Web services raises problem of becoming obsolete when changes occur on information sources. To avoid becoming obsolete, when information sources change their contents and/or their schema, Web services have to be substituted in order to ensure the integrity, the accessibility, the availability and the consistency of the afforded information. We consider that a Web Service is affected when one or more of its associated information sources are affected by schema changes. A critical challenge therefore is to design a system able to substitute affected Web services. In our solution we aim to propose a mediator able to integrate information sources into Web services while addressing the synchronization issue for affected Web services based on EVE framework [1]. Since EVE system proposes a prototype solution to automate view definitions rewriting thanks to Meta knowledge about information space formed by information sources, to Meta knowledge about user space constituted by evolving view definitions, and view synchronization algorithms [2][3].

We propose to take advantages of this approach and to adapt it in our context which is Web services. Our system revolves around three main components which are:

- A Web services Meta Knowledge Base WSMKB containing Web services, information sources and substitution constraints.

- A Web services View Knowledge Base WSVKB containing Web services and related views definition.

- Web services synchronization algorithm AS²W substituting affected Web services after schema changes using WSMKB and WSVKB constraints.

As illustration, we adopted a case study related to a domain characterized by a prominent need for information integrity and availability: Healthcare services.

This paper is organized as follows: Section 2 describes the related .Section 3 introduces the Web service model for gathering information sources. In section 4, we present the Web services synchronization solution by presenting the middleware main components which are the WSMKB, the WSVKB and the Web services synchronization algorithm and our illustration by the healthcare application. In section 5, we introduce our Web services synchronization algorithm AS2W. And section 6 concludes our work and presents some insights for future work.

## 2. RELATED WORKS

In the information space, data providers are autonomous. However , they usually have control over the schemas of their information sources which raises the question of the influence of schema changes, that can render affected view definition undefined [4][5][6].   Different approaches for addressing this problem have been presented in the literature. Service synchronization or substitution based on the functional properties of components has been addressed by many authors [7, 8, 9, 10, 11].  What sets us apart from the proposed approaches is that we aim at addressing the service synchronization problem taking into account the detection of changes which can occur on information sources from which Web services are constructed. In this context, EVE project [12][13],  offers a generic framework within which a view adaptation is solved when underlying information sources change their capabilities.  It neither relies on a globally fixed domain nor on ontology of permitted classes of data, both strong assumptions that are often not realistic. Instead, views are built in the traditional way over a number of base schemas and those views are adapted to base schema changes by rewriting them using information space redundancy and relaxable view queries [14]. The benefit of this approach is that no predefined domain is necessary, and a view can adapt to changes in the underlying data by automatically rewriting user queries, thanks to synchronization algorithms.  This framework has opened up a new direction of research by identifying view synchronization as unexplored problem of current view technology in the WWW. Our approach distinguishes itself from EVE [12] by the fact that we rely on specific advanced applications on the WWW which are Web services. Another novelty of our approach   is to apply our work is the health care domain.

## 3. WEB SERVICES MODEL

In today's collaborative environment, Web services appear to be a privileged mean to interconnect applications across organizations. Web Services are software systems designed to support interoperable machine-to-machine interaction over a network [15]. They are modular applications with interface descriptions that can be published, located, and invoked across the Web [16].

Different formalisms are proposed in the literature for modelling Web services. In [17, 18, and 19], state machine formalism is used for the description of Web services. This choice is justified by the fact that the state machine is simple especially to describe Web services conversation. The states represent phases through it passes the service while interacting. The states are labelled with logic names and the transitions are labeled with operations. In [20, 21], Web services are modelled using state chart diagram which is a graphic representation of a state machine. The service chart diagram is based on the UML state chart diagram to specify Web services components. None of the studied formalisms can be suitable for modelling the changes that can occur on Web services. In this section we introduce a novel approach for modeling and specifying Web services.

This approach sheds the light on two types of behaviours which are presentation and dynamic parts where:

- The dynamic Web service includes information sources access using views
- The static Web service part contains the presentation components

Web service presentation and dynamic parts are executed iteratively as given in Figure 1

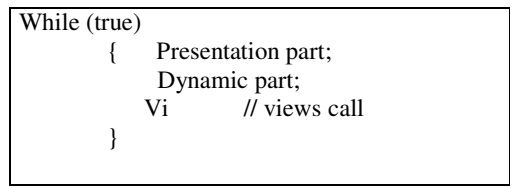

Figure 1. Web services model

Web services are constructed from views which are built from distributed, heterogeneous and autonomous information sources. Each information source has its own schema composed of relations and attributes. In several cases, Web service is undefined so it should be substituted by another Web service as modelled in figure 2.

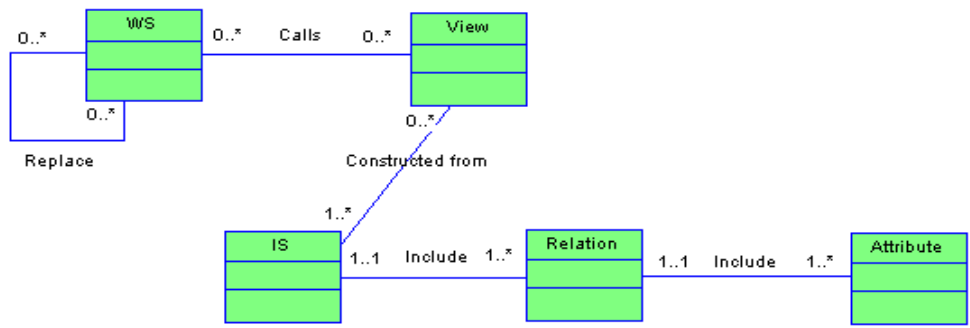

Figure 2 . Web service components relation

Types of relations between the different components are formalized in table 1

| |
|---|
| Let WS be a Web service, WS = {V1 ,…,Vn}<br>With Vi: views called by Web service WS,<br>      \|Vi\| ≥ 1,<br>      n : total number of views called by WS. |
| Let V be a view, V = {IS1 ,…, ISn}<br>With ISi : information sources from which the view V is constructed.<br>      \|ISi\| ≥ 1,<br>        n : total number of information sources from which the view V is constructed. |
| Let SI be an information source, IS = {R1 ,…, Rn}<br>With Ri : relations which belong to the information source IS,<br>      \|Ri\| ≥ 1,<br>      n : total number of relations which belong to the information source IS. |
| Let R be a relation, R = {A1 ,…, An}<br>With Ai : the attributes which belong to the relation R,<br>      \|Ai\| ≥ 1,<br>      n : total number of attributes which belong to the relation R. |
| Let WS be a Web service, WS= {WS1 ,…, WSn}<br>with WSi the Web service replacement list. |

Table 1. Relationship types between Web services components

In several cases, Web services are unavailable so we need to substitute them. In our case Web services are undefined due to schema changes which may render views (dynamic part) undefined. So Web service substitution reach on substituting Web Service dynamic part by rewriting affected views.

Let WS be a Web service and Vi the set of views defined accessed by Web service dynamic part. We suppose that the view V is undefined after schema changes. The Web service WS is synchronized to the Web service WS' with V rewritten on V'∈ Vi as shown in figure 3.

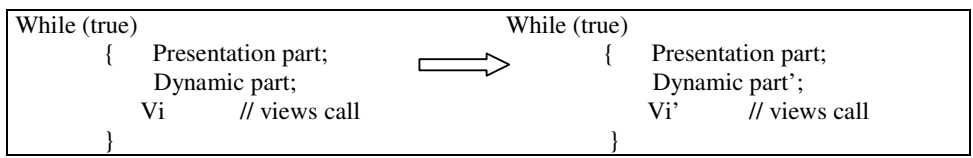

Figure 3 . Web service synchronization

The substitute Web service can be equivalent (≡), superset (⊇), subset (⊆) or indifferent (≈) to the initial Web service.

- The substitute Web service is equivalent (≡) to the initial Web service, if all dynamic part views of the substitute Web service are equivalent to all dynamic part views of the initial Web service.

- ▪ The substitute Web service is a superset ($\supseteq$) of the initial Web service, if at least one of the dynamic part views of the substitute Web service is a superset of one of the dynamic part views of the initial Web service.
- ▪ The substitute Web service is a subset ($\subseteq$) of the initial Web service, if at least one of the dynamic part views of the substitute Web service is a subset of one of the dynamic part views of the initial Web service.
- ▪ The substitute Web service is indifferent ($\approx$) of the initial Web service, if all dynamic part views of the substitute Web service are indifferent to all dynamic part views of the initial Web service.

## 4. WEB SERVICES SYNCHRONIZATION FRAMEWORK

Web services are built from distributed, heterogeneous, autonomous information sources which change continuously not only contents but also their schema which may render Web services undefined. We propose therefore a synchronization process which consists on detecting schema changes and substituting affected Web services. Only the two operations attribute deletion and relation deletion affect Web services. The Web service synchronization algorithm searches possible substitution of the affected component (attribute or relation) using WSMKB constraints and WSVKB constraints.

Our solution takes the form of a middleware connecting Web services to information sources as shown in figure 3 and is composed by:

- ▪ A Web services Meta Knowledge Base WSMKB containing Web services, information sources and substitution constraints
- ▪ A Web services View Knowledge Base WSVKB containing Web services and related views definition.
- ▪ Web services synchronization algorithm AS²W substituting affected Web services after schema changes using WSMKB and WSVKB constraints.

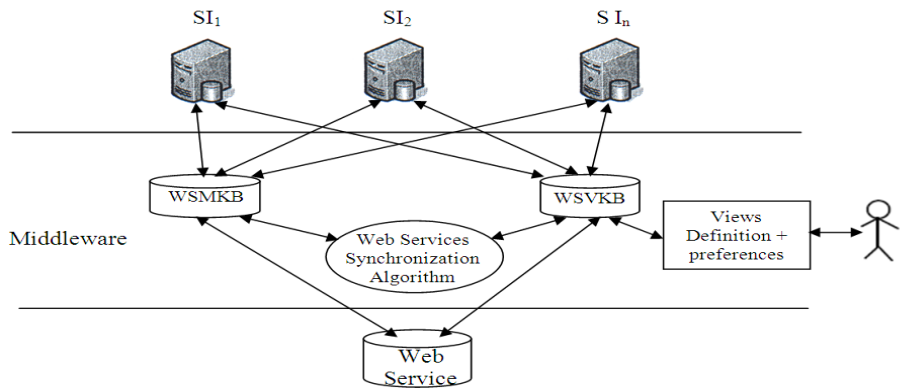

Figure 4. The system architecture

## 4.1. Web services Meta Knowledge Base (WSMKB)

Web services Meta Knowledge Base WSMKB contains information sources description as given in Figure 5; information sources joining the system must provide its structures and its contents to be stored in the WSMKB. Relationships between information sources have to be added to WSMKB as substitution rules as given in Figure 6, Figure 7 and Figure 8. The WSMKB constraints are represented respecting a model called MISD [22, 23]. WSMKB can be organized as follow:

- WS (WSid, WSISidList): Web services with information sources from which they are built as given in Figure 9.
- IS (ISid, ISRidList): information sources and their included relations. Relations (Rid, AttList): relations and their included attributes.
- The relationships between information sources or substitution constraints such as type integrity constraints, join constraints and partial/complete constraints.
- Replacement (WSid, WSreplacementList): Web services and their substitution Web services list as given in Figure 10.

In the following, we give an example of healthcare application. Each information sources have their own schemas and contents.

| S1 | Patient (IdP, Name, Age, Tel, IdH) |
| | Doctor (IdD, Name, Speciality) |
| | Hospital (IdH, Name, Localization) |
| | Doctor_Hospital (IdD, IdH) |
| | Diagnostic (IdP, IdD, DateT, Result) |
| | Operation (IdP, IdD, DateO, Result) |
| S2 | Patient (IdP, Name, Age, Tel, IdH, Med_Resp) |
| | Doctor (IdD, Name, Speciality, IdS) |
| | Hospital (IdH, Name, Localization) |
| | Doctor_Hospital (IdD, IdH) |
| | Diagnostic (IdP, IdD, DateT, Result) |
| | Operation (IdP, IdD, DateO, Result) |
| | Service (IdS, Speciality) |
| S3 | Patient (IdP, Name, Age, Tel) |
| | Doctor (IdD, Name, Speciality, Hospital, IdS) |
| | Hospital (IdH, Name, Localization) |
| | Patient_Hospital (IdP, IdH, IdD) |
| | Diagnostic (IdP, IdD, DateT, Result) |
| | Operation (IdP, IdD, DateO, Result) |
| | Service (IdS, Speciality) |

Figure 5. Information sources schemas.

A type integrity constraint of a relation $R(A1,…,An)$ states that an attribute $Ai$ is of domain type $Typei$. It allows verifying substitution possibility of an attribute by another while synchronizing Web services. A type integrity constraint is defined as follow:

$$TCR(A1,…,An) = R(A1,…,An) \subseteq A1(Type1) \times … \times An(Typen)$$

The type integrity constraints are expressed in the following

| TC | Type integrity constraints |
|---|---|
| TC1 | TCS1.Patient(IdP, Name, Age, Tel, IdH) = Patient (IdP, Name, Age, Tel, IdH) $\subseteq$ IdP(Number) $\times$Name(String)$\times$Age(Number)$\times$Tel(Number)$\times$IdH(Number) |
| TC2 | TCS1.Doctor(IdD, Name, Speciality) = Doctor (IdD, Name, Speciality) $\subseteq$ IdD(Number) $\times$Name(String) $\times$ Speciality(String) |
| TC3 | TCS1.Hospital (IdH, Name, Localization) = Hospital (IdH, Name, Localization) $\subseteq$ IdH(Number) $\times$Name(String) $\times$Localization(String) |
| TC4 | TCS1.Doctor_Hospital (IdD, IdH)=Doctor_Hospital (IdD, IdH) $\subseteq$ IdD(Number)$\times$IdH(Number) |
| TC5 | TCS1.Diagnostic (IdP, IdD, DateT, Result) = Diagnostic (IdP, IdD, DateT, Result)$\subseteq$ IdP(Number)$\times$IdD(Number)$\times$DateT(Date)$\times$Result(String) |
| TC6 | TCS1.Operation(IdP, IdD, DateO, Result)=Operation(IdP, IdD, DateO, Result) $\subseteq$ IdP(Number) $\times$IdD(Number)$\times$DateO(Date)$\times$Result(String) |
| TC7 | TCS2.Patient(IdP, Name, Age, Tel, IdH, Med_Resp) = Patient(IdP, Name, Age, Tel, IdH, Med_Resp) $\subseteq$IdP(Number)$\times$Name(String)$\times$Age(Number)$\times$Tel(Number)$\times$IdH(Number) $\times$Med_Resp(Number) |
| TC8 | TCS2.Doctor(IdD, Name, Speciality, IdS) = Doctor(IdD, Name, Speciality, IdS) $\subseteq$ IdD(Number) $\times$Name(String)$\times$Speciality(String)$\times$IdS(Number) |
| TC9 | TCS2.Hospital (IdH, Name, Localization) = Hospital(IdH, Name, Localization) $\subseteq$ IdH(Number) $\times$Name(String)$\times$Localization(String) |
| TC10 | TCS2.Doctor_Hospital(IdD, IdH)=Doctor_Hospital (IdD, IdH) $\subseteq$ IdD(Number) $\times$IdH(Number) |
| TC11 | TCS2.Diagnostic (IdP, IdD, DateT, Result) = Diagnostic (IdP, IdD, DateT, Result) $\subseteq$ IdP(Number) $\times$IdD(Number)$\times$DateT(Date)$\times$Result(String) |
| TC12 | TCS2.Operation (IdP, IdD, DateO, Result) = Operation(IdP, IdD, DateO, Result) $\subseteq$ IdP(Number) $\times$IdD(Number)$\times$DateO(Date)$\times$Result(String) |
| TC13 | TCS2.Service(IdS, Speciality) = Service (IdS, Speciality) $\subseteq$ IdS(Number) $\times$ Speciality(String) |
| TC14 | TCS3.Patient(IdP, Name, Age, Tel) = Patient (IdP, Name, Age, Tel) $\subseteq$ IdP(Number) $\times$Name(String) $\times$Age(Number)$\times$Tel(Number) |
| TC15 | TCS3.Doctor(IdD, Name, Speciality, Hospital, IdS) = Doctor (IdD, Name, Speciality, Hospital, IdS) $\subseteq$ IdD(Number) $\times$Name(String)$\times$Speciality(String)$\times$Hospital(Number)$\times$IdS(Number) |
| TC16 | TCS3.Hospital (IdH, Name, Localization) = Hospital (IdH, Name, Localization) $\subseteq$ IdH(Number) $\times$Name(String)$\times$Localization(String) |
| TC17 | TCS3.Patient_Hospital (IdP, IdH, IdD) = Patient_Hospital (IdP, IdH, IdD) $\subseteq$ IdP(Number) $\times$IdH(Number)$\times$IdD(Number) |
| TC18 | TCS3.Diagnostic(IdP, IdD, DateT, Result) = Diagnostic (IdP, IdD, DateT, Result) $\subseteq$ IdP(Number) $\times$IdD(Number)$\times$DateT(Date)$\times$Result(String) |
| TC19 | TCS3.Operation (IdP, IdD, DateO, Result) = Operation (IdP, IdD, DateO, Result) $\subseteq$ IdP(Number)$\times$IdD(Number)$\times$DateO(Date)$\times$Result(String) |
| TC20 | TCS3.Service(IdS, Speciality) = Service (IdS, Speciality) $\subseteq$ IdS(Number)$\times$Speciality(String) |

Figure 6. Type integrity constraints.

Join constraint between two relations R1 and R2 states that attributes in R1 and R2 can be joined. It allows verifying substitution possibility of a relation by another while synchronizing Web services. Join constraint between two relations R1 and R2 is defined as follow: $JC_{R1,R2} = (C_1$ AND …AND $C_n)$ In figure 7 we state the list of the join constraints related to our example

| TC | Join constraints |
|---|---|
| JC1 | S1.Patient.Name = S2.Patient.Name |
| JC2 | S1.Patient.Name = S3.Patient.Name |
| JC3 | S1.Doctor.Name = S2.Doctor.Name |
| JC4 | S1.Doctor.Name = S3.Doctor.Name |
| JC5 | S1.Doctor.Speciality = S2.Doctor.Speciality |
| JC6 | S1.Doctor.Speciality = S3.Doctor.Speciality |
| JC7 | S1.Hospital.Name = S2.Hospital.Name |
| JC8 | S1.Hospital.Name = S3.Hospital.Name |
| JC9 | S1.Hospital.Localization = S2.Hospital.Localization |
| JC10 | S1.Hospital.Localization = S3.Hospital.Localization |
| JC11 | S2.Service.Speciality = S3.Service. Speciality |

Figure 7. Join constraints.

Partial/complete constraint between two relations R1 and R2 states that the relation R1 (or a fragment of the relation R1) is a subset, a superset or equivalent to the relation R2 (or a fragment of the relation R2). Partial/complete constraint is defined as follow:

$$PC_{R1,R2} = (\pi_{Ai1,...,Aik}(\sigma_{C(Aj1,...,Ajp)}R1) \; \theta \; \pi_{An1,...,Ank}(\sigma_{C(Am1,...,Aml)}R2))$$

| TC | Partial/ complete constraints |
|---|---|
| PC1 | PCS1.Patient,S2.Patient = ($\pi$IdP, Name, Age, Tel(S1.Patient) $\subseteq$ $\pi$IdP, Name, Age, Tel(S2.Patient)) |
| PC2 | PCS1.Doctor,S2.Doctor = ($\pi$IdD, Name, Speciality(S1.Doctor) $\subseteq$ $\pi$IdD, Name, Speciality (S2.Doctor)) |
| PC3 | PCS1.Hospital,S2.Hospital=($\pi$IdH, Name, Localization(S1.Hospital)$\supseteq$ $\pi$IdH, Name, Localization (S2.Hospital)) |
| PC4 | PCS1.Operation,S2.Operation=$\pi$IdP, IdD, DateO, Result(S1.Operation) $\subseteq$ $\pi$IdP, IdD, DateO, Result (S2.Operation)) |
| PC5 | PCS2.Service,S3.Service = ($\pi$IdS, Speciality(S2.Service) $\supseteq$ $\pi$IdS, Speciality (S3.Service)) |

Figure 8. Partial/ complete constraints.

| |
|---|
| (WS1, {S1, S2}): The Web service WS1 is construct from information sources S1 and S2 <br> (WS2, {S1, S2}): The Web service WS2 is construct from information sources S1 and S2 <br> (WS3, {S3}) : The Web service WS3 is construct from information sources S3 |

Figure 9. Relation between Web services and information sources.

| |
|---|
| (WS1, {WS2, WS3}): The Web service WS1 can be replaced by the Web service WS2 or WS3 <br> (WS2, {WS3}): The Web service WS2 can be replaced by the Web service WS3 |

Figure 10. Web services substitution constraints.

## 4.2. Web services View Knowledge Base WSVKB

The Web Services View Knowledge Base WSVKB contains views definition using E-SQL and relations between Web services and its accessed views as given in Figure 12. E-SQL [22] language allows user preferences inclusion in views definition to indicate how views can evolve after schema changes.

E-SQL is an extension of SELECT-FROM-WHERE and respect the following syntax:

```
CREATE VIEW V [(column_list)] [VE= ['⊇' | '⊆' | '≡' | '≈'] AS
SELECT Attribute_Name [(AD = [true | false], AR = [true | false])]
        [, Attribute_Name [(AD = [true | false], AR = [true | false])]…]
FROM    Relation_Name [(RD = [true | false], RR = [true | false])]
        [, Relation_Name [(RD = [true | false], RR = [true | false])]…]
WHERE Primitive_Clause [(CD = [true | false], CR = [true | false])]
        [, Primitive  Clause [(CD = [true | false], CR = [true | false])]…];
```

Figure 11.  Structure of E-SQL query.

In an E-SQL query, each attribute, relation or condition has two evolution parameters. The dispensable parameter indicates if view components can be conserved (XD=False) or dropped (XD=True) from the substitute view. The replaceable parameter indicates if view components can be substituted (XR=True) or not (XR=False). Here X can be an attribute, a relation or a condition and the default value is False. View extension parameters VE proposed by E-SQL states that the substitute view can be equivalent (≡), superset (⊇), subset (⊆) or indifferent (≈) to the initial view.

WSVKB contents can be organized as follow:

▪ VIEW (VDid, VDText) : View definition using E-SQL.
▪ WS (WSid, VDidList) : Web services and their views definition list.

**Example 1**

We need to have doctors list from S1 having « Cardiologist » specialty, and accepting substitution of the relation S1.Doctor by the relation S2.Doctor, and accepting substitution of the attribute Name from the relation S1.Doctor by the attribute Name from the relation S2.Doctor.

```
CREATE VIEW      V1 VE='⊇' AS
SELECT           D.IdD, D.Name (AD=false, AR=true)
FROM             S1.Doctor D (RD=false, RR=true)
WHERE            (D.Speciality= "Cardiologist") (CD=false, CR=true);
```

**Example 2**

We need to have hospitals list from S1 localized in « Tunis » and accepting substitution of the relation S1.Hospital by the relation S2.Hospital, and accepting substitution of the attribute Name from the relation S1.Hospital by the attribute Name from the relation S2.Hospital.

```
CREATE VIEW      V2 VE='⊆' AS
SELECT           H.IdH, H.Name (AD=false, AR=true)
FROM             S1.Hospital H (RD=false, RR=true)
WHERE            (H.Localization= "Tunis") (CD=false, CR=true);
```

```
(WS1, {V1, V2, V3}): Web service WS1 is constructed from V1, V2 and V3.
(WS2, {V3, V4, V5}): Web service WS2 is constructed from V3, V4 and V5.
(WS3, {V6}): Web service WS3 is constructed from V6.
```

Figure 12. Relation between Web services and views.

### 4.3 Case Study

Web services synchronization consists on automatically rewriting or substituting Web services affected after schema changes referring to WSMKB constraints and to WSVKB constraints.

The synchronization process consists on detecting schema changes (relations or attributes deletion), detecting affected Web services and applying synchronization algorithm to determine possible substitution of the affected Web services.

**Case 1**

Suppose that Name attribute from the relation S1.Doctor is deleted, then it's substituted by Name attribute from the relation S2.Doctor since [TCS1.Doctor(IdD, Name, Speciality)=Doctor(IdD, Name, Speciality) $\subseteq$ IdD(Number) $\times$ Name(String) $\times$ Speciality(String)] and [TCS2.Doctor(IdD, Name, Speciality, IdS)= Doctor(IdD, Name, Speciality, IdS) $\subseteq$ IdD(Number) $\times$ Name(String) $\times$ Speciality(String) $\times$ IdS(Number)] and [PCS1.Doctor,S2.Doctor=($\pi$IdD, Name, Speciality(S1.Doctor) $\subseteq$ $\pi$IdD, Name, Speciality(S2.Doctor))] and [S1.Doctor.Name = S2.Doctor.Name]. The view definition of V1 becomes:

```
CREATE VIEW    V1' VE='⊇' AS
SELECT              D.IdD, D2.Name (AD=false, AR=true)
FROM                S1.Doctor D (RD=false, RR=true),
                    S2.Doctor D2 (RD=false, RR=true)
WHERE               (D.Speciality= "Cardiologist") (CD=false, CR=true) AND
            (D.IdD = D2.IdD
```

**Case 2**

Suppose that S1.Hospital relation is deleted, then it's substituted by S2.Hospital relation since [PCS1.Hospital,S2.Hospital = ($\pi$IdH, Name, Localization(S1.Hospital) $\supseteq$ $\pi$IdH, Name, Localization (S2.Hospital))]. The view definition of V2 becomes:

```
CREATE VIEW V2'      VE='⊆' AS
SELECT               H2.IdH, H2.Name (AD=false, AR=true)
FROM                 S2.Hospital H2 (RD=false, RR=true)
    WHERE                (H2.Localization= "Tunis") (CD=false, CR=true);
```

## 5. WEB SERVICES SYNCHRONIZATION ALGORITHMS

Web services are composed by presentation and dynamic parts including information sources access using views call. As previously said dynamic part includes services gathered from information sources, the latter change continuously which may render views undefined then may render Web services undefined and inaccessible. So these Web services must be substituted by other ones.

Web services synchronization consists on substituting affected Web services referring to constraints embodied into the WSMKB and into the WSVKB. So synchronization process consists on detecting change and according to this change Delete_Attribute procedure or Delete_Relation procedure will be executed as given in Algorithm 1. Only the two operators delete attribute and delete relation are treated by our algorithm.

Algorithm 1 Synchronization

```
00. BEGIN
01.      Input = {schema change};    /* changes can be an attribute deletion or a relation deletion */
02.      Output = {synchronized services};
03.      FOR each changes
04.              IF (Input = attribute deletion) THEN
05.                      Delete_Attribute (A);    /*A: deleted attribute*/
06.              ELSE
07.                      Delete_Relation (R);       /*R: deleted relation*/
08.              END IF
09.      END FOR
10. END
```

## 5.1. Relation deletion

The deletion of a relation R affects Web services if it appears in at least one of the views that Web services dynamic part references. To synchronize Web services affected after a relation deletion, we must verify if this relation is replaceable or not and if it's dispensable or not. So we must verify the evolution parameters; dispensable and replaceable parameters.

- If the relation is dispensable (RD = True) and not replaceable (RR = False) then this relation can be omitted from the substitute view, then from the substitute Web service.
- If the relation is dispensable (RD = True) and replaceable (RR = True) then it's substituted if there is a substitution relation, else it can be omitted from the substitute view, then from the substitute Web service.
- If the relation is indispensable (RD = False) and not replaceable (RR = False) then failure will be returned and the Web service can't be synchronized.
- If the relation R is indispensable (RD = False) and replaceable (RR = True) then if a substitution relation S exists, then R will be substituted by the relation S, else failure will be returned and the Web service can't be synchronized.

A relation S can substitute a relation R if all attributes of the relation R which are indispensable and replaceable (AD = False and AR = True) and appear in SELECT and WHERE clause have substitute attributes in the relation S, and have the same type.

Relation deletion affects a set of Web services, so executing Delete_Relation procedure, we have the affected one and Web services substitution will be done if it's possible.

Algorithm 2  PROCEDURE Delete_Relation (R)

```
00. BEGIN
01.      SA = SearchSA (R);  /* search effected Web services*/
02.      FOR each SA
03.              Search_Substitution (SA, R);  /* search Web services substitution */
04.      END FOR
05. END
```

As given in Algorithm 3, affected Web services are those who are referenced by the deleted relation. So SearchSA procedure will search affected Web services referring to constraints embodied into the WSMKB.

Algorithm 3 PROCEDURE SearchSA (R)

```
00. BEGIN
01.      /* search in WSMKB Web services referenced by the relation R deleted*/
02. END
```

Web services synchronization reach on views synchronization, this synchronization is done referring to preferences embodied into the WSVKB. So executing Serach_Substitution procedure as given in Algorithm 4 and according to relation evolution parameters a set of treatments will be executed.

Algorithm 4 PROCEDURE Search_Substitution (SA, R)

```
00. BEGIN
01.      ListViews = {views containing R and appear in SA};
02.      FOR each view V of ListViews
03.              IF (RD = TRUE and RR = FALSE) THEN
04.                      Delete R from V;
05.              ELSE IF (RD = TRUE and RR = TRUE) THEN
06.                      IF (Find _Relation (R, S)) THEN
07.                              Substitute (R, S);
08.                      ELSE
09.                              Delete R from V;
10.                      END IF
11.              ELSE IF (RD = FALSE AND RR = FALSE) THEN
12.                      Return failure with msg "Web service can't be synchronized";
13.              ELSE IF (RD = FALSE AND RR = TRUE) THEN
14.                      IF (Find_Relation (R, S)) THEN
15.                              Substitute (R, S);
16.                      ELSE
17.                              Return failure with msg "Web service can't be synchronized";
18.                      END IF
19.              END IF
20.      END FOR
21. END
```

As given in Algorithm 5, Find_Relation procedure will find in WSMKB a substitute relation to the deleted one. It will be substitution if it exists in the WSMKB a relation that substitute the deleted relation. So Replace procedure verifies if two relations are replaceable or not as shown in algorithm 6.

Algorithm 5 Boolean PROCEDURE Find_Relation (in: R, out: S)

```
00. BEGIN
01.      FOR each relation S
02.              IF Replace (R, S) THEN
03.                      ListRelation = ListRelation + S;
04.              END IF
05.      END FOR
06.      IF ListRelation = empty THEN
07.              Return (FALSE);
08.      ELSE
09.              S = {relation ∈ ListRelation which best substitute R};
10.              Return (TRUE);
11.      END IF
12. END
```

Algorithm 6 Boolean PROCEDURE Replace (R, S)

```
00. BEGIN
01.      ListAttributes = {attributes which appears in SELECT and WHERE clause and are
indispensables and replaceable};
02.      IF the attributes of S substitute ListAttributes THEN
03.              Return (TRUE);
04.      ELSE
05.              Return (FALSE);
06.      END IF
07. END
```

As given in Algorithm 7, executing Substitute procedure, will substitute the attributes of the deleted relation R that appears in the SELECT clause and the WHERE clause with the attributes of S and will substitute the deleted relation R with the relation S.

Algorithm 7 PROCEDURE Substitute (R, S)

```
00. BEGIN
01.      Replace the attributes of R with the attributes of S in the SELECT clause and the WHERE clause;
02.      Replace the relation R with S;
03. END
```

## 5.2. Attribute deletion

A deleted attribute A can affects Web services if it appears in at least one of the views that Web service dynamic part references. To synchronize Web service affected after an attribute deletion, we must verify if this attribute is replaceable or not and if it's dispensable or not. So we must verify the evolution parameters; dispensable and replaceable parameters.

If the attribute is dispensable (AD = True) and not replaceable (AR = False) then this attribute can be omitted from the substitute view, then from the substitute Web service.

If the attribute is dispensable (AD = True) and replaceable (AR = True) then it's substituted if there is a substitute attribute, else it can be omitted from the substitute view, then from the substitute Web service.

If the attribute is indispensable (AD = False) and not replaceable (AR = False) then failure will be returned and Web service can't be synchronized.

If the attribute is indispensable (AD = False) and replaceable (AR = True) then this attribute will be substituted if a substitute attribute exists, else failure will be returned and Web service can't be synchronized.

An attribute B from a relation S can substitute an attribute A from a relation R if these attributes have the same types in other words if exists in WSMKB a type integrity constraint such that:

TC (R.A) = R (A) $\subseteq$ A (type) and TC (S.B) = S (B) $\subseteq$ B (type) and exists in WSMKB a join constraint between the two attributes: JCR, S = (R.A = S.B)

Attribute deletion affects a set of Web services, so executing Delete_Attribute procedure, we have the affected one and Web services substitution will be done if it's possible.

Algorithm 8 PROCEDURE Delete_Attribute (A)

```
00. BEGIN
01.       SA = SearchSA (A);  /* search affected services */
02.       FOR each SA
03.              Search_Substitution (SA, A);  /* search Web services substitution */
04.       END FOR
05. END
```

As given in Algorithm 9, affected Web services are those who are referenced by the deleted attribute. So SearchSA procedure will search affected Web services referring to constraints embodied into the WSMKB.

Algorithm 9 PROCEDURE SearchSA (A)

```
00. BEGIN
01.       /* search in WSMKB Web services referenced by the deleted attribute A*/
02. END
```

Web services synchronization reach on views synchronization, this synchronization is done referring to preferences embodied into the WSVKB. So executing Serach_Substitution procedure as given in Algorithm 10 and according to attribute evolution parameters a set of treatments will be executed.

### Algorithm 10 PROCEDURE Search_Substitution (SA, A)

```
00. BEGIN
01.     ListViews = {views containing A and appear in SA};
02.     FOR each view V from ListViews
03.             IF (AD = TRUE AND AR = FALSE) THEN
04.                     Delete A from V;
05.             ELSE IF (AD = TRUE AND AR = TRUE) THEN
06.                     IF (Find_Attribute (A, B)) THEN
07.                             Substitute (A, B);
08.                     ELSE
09.                             Delete A from V;
10.                     END IF
11.             ELSE IF (AD = FALSE AND AR = FALSE) THEN
12.                     Return failure with msg "Web service can't be synchronized";
13.             ELSE IF (AD = FALSE AND AR = TRUE) THEN
14.                     IF (Search_Attribute (A, B)) THEN
15.                             Substitute (A, B);
16.                     ELSE
17.                             Return failure with msg "Web service can't be synchronized";
18.                     END IF
19.             END IF
20.     END FOR
21. END
```

As given in Algorithm 11, Find_Attribute procedure will find in WSMKB a substitute attribute to the deleted one. It will be substitution if it exists in the WSMKB an attribute that substitute the deleted attribute. So Replace procedure as shown in algorithm 12 verify if two attributes are replaceable or not.

### Algorithm 11 Boolean PROCEDURE Find_Attribute (in: A, out: B)

```
00. BEGIN
01.     For each attribute B
02.             IF Replace (A, B) THEN
03.                     ListAttributes = ListAttributes + B;
04.             END IF
05.     END FOR
06.     IF ListAttributes = empty THEN
07.             Return (FALSE);
08.     ELSE
09.             B = {attribute ∈ ListAttributes which best substitute A};
10.             Return (TRUE);
11.     END IF
12. END
```

Algorithm 12 Boolean PROCEDURE Replace (A, B)

```
00. BEGIN
01.     IF (TC(R.A)=R (A) ⊆ A(type) AND TC(S.B)=S(B) ⊆ B(type) AND JCR, S
=(R.A=S.B)) THEN
02.             Return (TRUE);
03.     ELSE
04.             Return (FALSE);
05.     END IF
06. END
```

As given in Algorithm 13, executing Substitute procedure, will substitute the deleted attribute A in the SELECT clause and/or the WHERE with B, will add the relation containing B to the FROM clause and will add the join constraint between the relation containing the deleted attribute A and the relation containing the attribute B.

Algorithm 13 PROCEDURE Substitute (A, B)

```
00. BEGIN
01.     IF A appears in SELECT clause THEN
02.             Delete A from SELECT clause;
03.             Add the relation S containing B to the FROM clause;
04.             Add join constraint between R and S;
05.             Add B to the SELECT clause;
06.     ELSE IF A appears in WHERE clause THEN
07.             C = {constraint containing A};
08.             IF CD=TRUE AND CR = FALSE THEN
09.                     Delete the constraint containing A from the WHERE clause;
10.             ELSE IF CD= FALSE AND CR = FALSE THEN
11.                     Return failure with msg "Web service can't be synchronized";
12.             ELSE
13.                     Delete constraint containing A from the WHERE clause;
14.                     Add the relation S containing B to the FROM clause;
15.                     Add join constraint between R and S;
16.                     Add the new constraint containing the new attribute B to the WHERE clause;
17.             END IF
18.     ELSE IF A appears in SELECT clause and in WHERE clause THEN
19.             Delete A from SELECT clause;
20.             Add the relation S containing B to the FROM clause;
```

```
21.             Add join constraint between R and S;
22.             Add B to the SELECT clause;
23.             IF CD = TRUE and CR = FALSE THEN
24.                     Delete the constraint containing A from the WHERE clause;
25.             ELSE IF CD = FALSE and CR = FALSE THEN
26.                     Return failure with msg "Web service can't be synchronized";
27.             ELSE
28.                     Delete the constraint containing A from the WHERE clause;
29.                     Add the new constraint containing the new attribute B to the WHERE clause;
30.             END IF
31.     END IF
32. END
```

## 6. IMPLEMENTATION

A prototype of the proposed system has been implemented. We used AXIS 1.1 which is Java platform for creating and deploying web services applications for creating Web services
The graphical user interface, the WSMKB, the WSVKB, and the view synchronizer are implemented using Java, and the participating ISs are built on Microsoft Access. The communication between the system and the information space is via JDBC. The view synchronization algorithms for the different basic schema changes presented in Section 7 have been implemented.

## 7. CONCLUSION

In this paper we proposed a solution to the problem of Web services synchronization caused by changes which can occur to information sources from which Web services are built and which may render Web services partially or totally inaccessible .
We have presented as solution a middleware connecting Web services to information sources. The middleware is composed by a Web service Meta Knowledge Base WSMKB, a Web Service View Knowledge WSVKB and Web services synchronization algorithms. Our model proved the feasibility of marrying Web services concepts, and the EVE approach [12] which offers a solid foundation for addressing the general problem of how to maintain views in dynamic environments.
Future work focus on a total synchronization of Web Services and will not be limited to the two operations attribute deletion and relation deletion which affect Web service. We also intend to develop algorithms for view maintenance of the view extent under both schema and data changes of the information sources.

## REFERENCES

[1]     Amy J. Lee,Anisoara Nica,A. Rundensteiner, "The EVE Approach: View Synchronization in Dynamic Distributed Environments",IEEE Transactions on Knowledge and Data Engineering, 2002,pp.931-954.

[2]     X. Zhang, E. A. Rundensteiner, L. Ding, "PVM: Parallel View Maintenance Under Concurrent Data Updates of Distributed Sources, in Data Warehousing and Knowledge Discovery",Proceedings, 2001,pp. 230–239.

[3]     A. J. Lee, A. Nica, and E. A. Rundensteiner, "The EVE Framework: View Evolution in an Evolving Environment", Technical Report WPI-CS-TR-97-4, Worcester Polytechnic Institute, Dept. of Computer Science, 1997.

[4]     J. A. Blakeley, P.-E. Larson, and F. W. Tompa. "Efficiently Updating Materialized Views". Proceedings of SIGMOD, 1986, pp. 61–71.

[5]     Y. Zhuge, H. Garc´ıa-Molina, and J. L. Wiener, "The Strobe Algorithms for Multi-SourceWarehouse Consistency",In International Conference on Parallel and Distributed Information Systems, 1996 pp. 146–157.

[6]     D. Agrawal, A. El Abbadi, A. Singh, and T. Yurek. "Efficient View Maintenance at Data Warehouses",In Proceedings of SIGMOD, 1997, pp. 417–427.

[7]     B. Benatallah, F. Casati, and F. Toumani, " Representing,analysing and managing web service protocols".,Data Knowl.Eng, 2006, pp. 327–357.

[8]     L. Bordeaux, G. Salaun, D. Berardi, and M. Mecella, "When are two web services compatible" Lecture Notes in Computer Science, 2005, pp. 15–28.

[9]     F. Liu, L. Zhang, Y. Shi, L. Lin, and B. Shi, "Formal analysis of compatibility of web services via ccs". In Proc. of theInternational Conference on Next Generation Web Services Practices, IEEE Computer Society, 2005, pp. 143.

[10]    A. Martens, S. Moser, A. Gerhardt, and K. Funk, " Analyzing compatibility of bpel processes". In International Conference on Internet and Web Applications and Services, 2006.

[11]    J. Pathak, S. Basu, and V. Honavar, " On context-specific substitutability of web services" In Proc. of the International Conference on Web Services,. IEEE Computer Society, 2007, pp. 192– 199.

[12]    E. A. Rundensteiner, A. J. Lee, and A. Nica," On PreservingViews in Evolving Environments",In Proceedings of 4th Int, Athens, Greece,1997, pp.13.1–13.11.

[13]    A. Koeller and E. A. Rundensteiner," History-Driven View Synchronization", Springer Verlag , Greenwich, UK,2000, pp.  168–177.

[14]    A. Nica,"View Evolution Support for Information Integration Systems over Dynamic Distributed Information Spaces",PhD thesis, University of Michigan in Ann Arbor, in progress 1999.

[15]    W3C. Web Services Architecture,  URL: http://www.w3.org/TR/ws-arch/,2004.

[16]    Jean Dollimore, Tim Kindberg, and George Coulouris. "Distributed Systems  Concepts and Design". Addison Wesley/Pearson Education, 4th edition, 2005.

[17]    Boualem Benatallah, Fabio Casati, Farouk Toumani. "Web Service Conversation Modeling". Published by the IEEE Computer Society ,2004.

[18]    Boualem Benatallah, Fabio Casati, Farouk Toumani, and Rachid Hamadi. "Conceptual Modeling of Web Service Conversations". CAiSE ,2003

[19]    Julien Ponge. "Modeling and Analyzing Web Services Protocols",2005.

[20]    Djamal Benslimane, Zakaria Maamar, Chirine Ghedira. "A View-based Approach for Tracking Composite Web Services". Proceedings of the Third European Conference on Web Services (ECOWS'05). 2005.

[21]    Z. Maamar, B. Benatallah, and W. Mansoor. "Service Chart Diagrams - Description & Application". In Proceedings of The Alternate Tracks of The 12th International World Wide Web Conference (WWW'2003), 2003.

[22]    E. a. Rundensteiner, A. J. Lee and A. Nica. "The EVE Framework: View Evolution in an Evolving Environment". Technical Report WPI-CS-TR-97-4. December 1997.

[23]    Anisora Nica, Amy J.Lee et Elke A. Rundensteiner. "The CVS Algorithm for View Synchronisation in Evolvable Large-Scale Information System". Technical Report WPI-CS-TR-97-8. September 1997.